\documentclass[10pt,twocolumn,letterpaper]{article}

\usepackage{cvpr}              

\usepackage[dvipsnames]{xcolor}

\definecolor{cvprblue}{rgb}{0.21,0.49,0.74}
\usepackage[pagebackref,breaklinks,colorlinks,citecolor=cvprblue]{hyperref}
\usepackage{bbm}

\def\paperID{*****} 
\def\confName{CVPR}
\def\confYear{2024}

\title{MMIST-ccRCC: A Real World Medical Dataset for the Development of Multi-Modal Systems}

\author{Tiago Mota$^{1}$, M. Rita Verdelho$^{1}$, Diogo J. Ara\'ujo$^{1}$, Alceu Bissoto$^{2}$, Carlos Santiago$^{1}$, Catarina Barata$^{1,3}$\\
$^{1}$Institute for Systems and Robotics, LARSyS, Instituto Superior T\'ecnico, Portugal\\
$^{2}$Institute of Computing, Recod.ai Lab, University of Campinas, Brazil\\
$^{3}$Lisbon ELLIS Unit \\
{\tt\small ana.c.fidalgo.barata@tecnico.ulisboa.pt}
}

\begin{document}
\maketitle
\begin{abstract}
The acquisition of different data modalities can enhance our knowledge and understanding of various diseases, paving the way for a more personalized healthcare. Thus, medicine is progressively moving towards the generation of massive amounts of multi-modal data (\emph{e.g,} molecular, radiology, and histopathology). While this may seem like an ideal environment to capitalize data-centric machine learning approaches, most methods still focus on exploring a single or a pair of modalities due to a variety of reasons: i) lack of ready to use curated datasets; ii) difficulty in identifying the best multi-modal fusion strategy; and iii) missing modalities across patients. In this paper we introduce a real world multi-modal dataset called MMIST-CCRCC that comprises 2 radiology modalities (CT and MRI), histopathology, genomics, and clinical data from 618 patients with clear cell renal cell carcinoma (ccRCC). We provide single and multi-modal (early and late fusion) benchmarks in the task of 12-month survival prediction in the challenging scenario of one or more missing modalities for each patient, with missing rates that range from 26$\%$ for genomics data to more than 90$\%$ for MRI. We show that even with such severe missing rates the fusion of modalities leads to improvements in the survival forecasting. Additionally, incorporating a strategy to generate the latent representations of the missing modalities given the available ones further improves the performance, highlighting a potential complementarity across modalities. Our dataset and code are available here: \href{https://multi-modal-ist.github.io/datasets/ccRCC}{multi-modal-ist.github.io/datasets/ccRCC}.           
\end{abstract}    
\section{Introduction}
\label{sec:intro}
Clinical practice is steadily advancing towards a more patient-centric vision, moving from the ``one-size-fits-all'' approach to more personalized treatments \cite{steyaert2023multimodal}. 
An example is the field of oncology, where patient management is currently determined through a combination different sources of information: genomics and proteomics (molecular) data are used as predictive biomarkers of drug response and risk stratification; histopathology allows the inspection of the tumour at the cellular level, conveying details about its spatial heterogeneity; and radiology, namely computerized tomography (CT) and magnetic resonance imaging (MRI), which allows the inspection of the tumour's 3D structure and the overall tissue morphology. 

It is clear that a single modality is not sufficient to capture all the information about a patient. Thus, there is an increased demand for tools that can integrate various modalities and find synergies between them. This has fostered the research on novel multi-modal approaches to address various oncology problems, such as tumour subtyping \cite{furbock2022} or patient prognosis \cite{esteva2022,zuo2022,cui2022}. 

Nevertheless, the success of multi-modal approaches is limited by a set of challenges \cite{steyaert2023multimodal}. 
The curation of a multi-modal dataset is a difficult process, hampered by not only the high volume and dimension of the data, but also by the time-consuming task of compiling data that comes from different sources. The different modalities are often not all available for every patient, leading to challenging missing data scenarios. And while several clinical centers are now digitizing all the patient information, it may be spread across various devices. Moreover, in order to be useful to train a model, the data must be annotated by experts, which is time consuming and often infeasible at a large scale. 
Consequently, most of the research is still conducted on single or pair of modalities.

Public repositories such as The Cancer Genome Atlas (TCGA) or the The Cancer Imaging Archive (TCIA) \cite{tcia} are becoming increasingly popular \cite{tcga21,tcga22,tcga23} due to the massive amount of data they have compiled. 
However, for most of the available multi-modal studies, data lacks curation and is spread across several repositories, limiting its usage to a subset of the modalities (\textit{e.g.}, histopathology and genomics~\cite{ning2021relation}). Additionally, a standardized dataset with training and test partitions is still missing for works that use the TCGA and TCIA datasets, making it hard to correctly benchmark the approaches that are being proposed. 

In this work, we propose to address the above challenges by publicly releasing a multi-modal dataset of 618 patients with clear cell renal cell carcinoma (ccRCC). This dataset, called MMIST-ccRCC, was curated from TCGA, TCIA, and the Clinical Proteomic Tumor Analysis Consortium (CPTAC), and comprises 2 radiology modalities (CT and MRI), histopathology, genomics, and clinical data. MMIST-ccRCC represents a real world scenario, where different modalities are missing across patients and it is also possible to have more than one CT/MRI scan. 

To foster further research, we provide a preliminary set of single and multi-modal benchmarks evaluated on MMIST-ccRCC for predicting patient survival. Both early and late fusion~\cite{erly-late} have been explored in this context. To deal with the multiple scans for CT and MRI, we propose a strategy based on multiple instance learning (MIL) to select the most suitable scans for each patient. Finally, to cope with the missing modalities, we adopt a generative method, which generates latent feature vectors of the missing modalities. By generating latent representations instead of the original modality, we bypass the bigger challenge of image synthesis. Our experimental results demonstrate that, while latent vectors are not the original modality, it is still possible to improve the results of the multi-modal benchmarks. 

Summarizing, this paper contributions are the following: 
\begin{enumerate}[label=\roman*)]
    \item The release of a curated real world multi-modal dataset called MMIST-ccRCC that comprises 2 radiology modalities (CT and MRI), histopathology, genomics, and clinical data for 618 patients. This dataset contains variable degrees of missing data across modalities, as will be discussed in Section \ref{sec:dataset}.
    
    \item The proposal of a set of benchmark approaches to assess the performance of single and multi-modal strategies on MMIST-ccRCC for predicting patient survival at 12-months. In particular, we compare several early and late fusion strategies and develop an innovative strategy based on MIL to select the most suitable CT and MRI scans per patient. Details about the benchmarks are given in Section \ref{sec:benchmarks} and their performance assessed in Section \ref{sec:exp_res}.
    
    \item The development and assessment of a generative approach to handle missing modalities. Our method bypasses the issue of generating the original modalities by instead inferring their latent vectors. This will be discussed in Section \ref{sec:benchmarks} and assessed in Section \ref{sec:exp_res}.
\end{enumerate}

\section{MMIST-ccRCC Dataset}
\label{sec:dataset}
ccRCC is the most common type of kidney cancer, making up to 80\% of all renal cell carcinoma cases in adults. Estimating the prognosis is critical for patient management, but is still a very challenging task~\cite{hotker2016clear}. The ongoing research on this topic has led to the creation of two public studies: CPTAC-CCRCC~\cite{CPTAC} and TCGA-KIRC~\cite{KIRC}, from which we curated MMIST-ccRCC. 

We collected data from patients with a follow-up of 12-months, resulting in a total of 189 patients for CPTAC-CCRCC and 429 patients for TCGA-KIRC. Both studies contain: radiology data -- CT and MRI; histopathology data -- whole slide images (WSI); clinical, and genomics data. In total, MMIST-ccRCC comprises 618 patients, of which 543 survived after 12 months ($88\%$). These patients were randomly split into training (80\%) and test (20\%) partitions using a stratified approach. Regarding the modalities, all patients present WSIs and clinical data, while genomics is available for 74\% of the patients, CT for 40\%, and MRI for 8\%. The final distribution of available modalities for each set is shown in Table \ref{tab:data_distribution}.

\begin{table}[t]
{\small
\begin{tabular}{@{}lcccccc@{}}
\cmidrule(l){2-7}
      & Patients & CT  & MRI & WSI & Genomics & Clinical \\ \cmidrule(l){2-7} 
Train & 497      & 189 & 35  & 497 & 361     & 497      \\ \midrule
Test  & 121      & 59  & 13  & 121 & 101     & 121      \\ \midrule
Total & 618      & 248 & 48  & 618 & 462     & 618      \\ \bottomrule
\end{tabular}}
\caption{Number of modalities per patient across the dataset.}\label{tab:data_distribution}
\end{table}

In the following sections we provide specific details regarding the curation process of each modality. For simplicity and given their characteristics, we group in one section the clinical and genomics modalities and in another all the imaging data. It is important to stress that the curation process was not trivial, since it was necessary to merge information from two different studies, and entailed the linking of patients across three data repositories: TCGA, CPTAC, and TCIA.

\subsection{Clinical and Genomics Data}
\label{sec:cligen_set}
Clinical data comprises information about the patient and the tumour, consisting primarily of numerical variables, with regular instances of categorical and ordinal types of data. There were some differences across the studies, regarding the amount, type, and naming of the variables. Additionally, there were variables with missing data above 50\%. Thus, we performed a variable selection, identifying those that should be merged between CPTAC-CCRCC and TCGA-KIRC. 

Our choice was mainly supported by the American Joint Committee on Cancer (AJCC) staging system\footnote{{\scriptsize https://www.facs.org/quality-programs/cancer-programs/american-joint-committee-on-cancer/cancer-staging-systems/}}. AJCC considers four relevant categories: i) the extent of the tumor (T); ii) the extent of spread to the lymph nodes (N); iii) the presence of distant metastasis (M); and iv) the assessment of M but based on pathological data (Mp). T is graded in 11 possible levels, while the other categories are all ranked between 0 and 2. The AJCC tumour staging was also considered (4 levels), as well as the tumour grading as assessed by the pathologist (1 to 5). Finally, patient metadada (age, gender, race, and cancer history) was included. Gender (male/female) and cancer history (yes/no) are binary variables, age was divided in intervals of 10 years, and race was normalized by the least detailed of the studies: ``Black or African American’', ``White'', ``Asian'', ``Hispanic or Latino'', or ``Other''. In the end, the clinical data comprised 11 variables.

Regarding the genomics features, we followed the study by \citet{genetic_mutations_CCRCC}, who identified key genetic mutations that characterize the genomics profile of ccRCC. Based on \cite{genetic_mutations_CCRCC}, we considered the following genes as relevant to be included in MMIST-ccRCC: VHL, PBMR1, and TTN. For each we assess whether a mutation is present or absent.

\subsection{Imaging Modalities}
\label{sec:img_set}
WSIs are digital representations of the tissue samples observed by the pathologist. In the case of MMIST-ccRCC, each patient is associated with at least two WSIs: the one used for diagnosing and staging the tumour and one additional WSI to characterize the tissue that surrounds the tumour and is assessed after the excision. Nevertheless, in some cases the number of WSIs is higher, leading to a final set of 2,573 images.

Regarding radiology data, the original datasets contain more than 2,200 CT and 1,000 MRI volumes. When compared with the number of patients that have at least one of the modalities (recall Table~\ref{tab:data_distribution}) it is clear that the number of scans per patient is very high. Thus, we have filtered out volumes that were deemed less relevant for the ccRCC problem. In particular, we discarded the localization scans, the pre-contrast ones, and scans that were acquired with a significant time lapse from the diagnosis (years). Moreover, we have also observed that the axial view was the most prevalent both in CT and MRI. This has led to the removal of scans from the coronal and sagittal views, to minimize domain shifts. In the end, we reduced the number of volumes to 736 for CT and 552 for MRI.
\begin{figure*}
  \centering
\includegraphics[width=.83\linewidth]{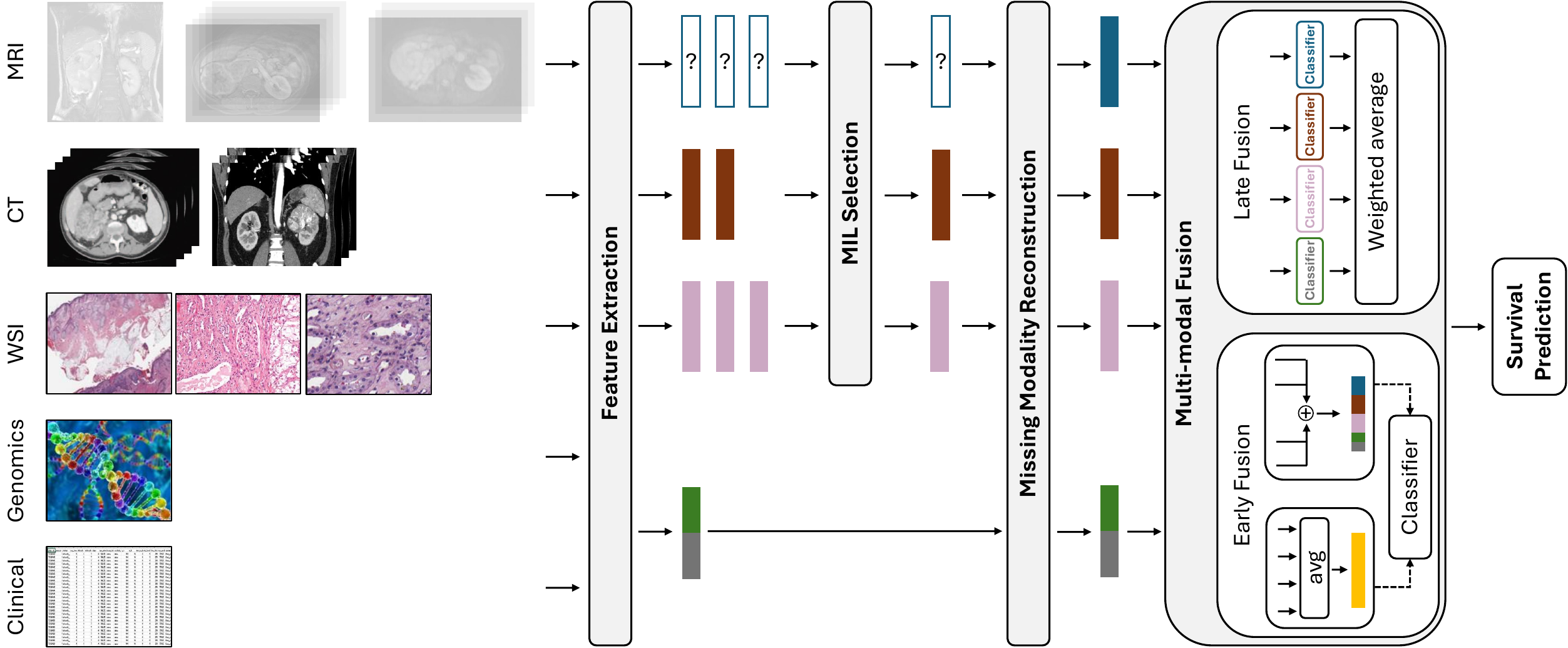}
  \caption{Overview of the proposed benchmarks. The vertical colored bars are the latent representation for each modality. In the example, MRI is the missing modality and is recovered in the reconstruction block.}
  \label{fig:overview}
\end{figure*}

\section{Benchmarks}
\label{sec:benchmarks}
MMIST-ccRCC is a rich dataset, which was curated to be used in a variety of medical tasks, such as biomarker prediction using imaging data \cite{esteva2022}, survival prediction \cite{cui2022}, modality generation to handle missing data \cite{missingdata-1}, and building relational multi-modal spaces~\cite{ning2021relation}.  
In this work, we perform a set of benchmarking experiments in the task of 12-month survival prediction, as this is a relevant and popular task within the multi-modal research field \cite{survival22-2}. Additionally, we investigate a generative approach to handle the missing modality problem.

We start by separately evaluating each of the modalities, followed by experiments on the multi-modal setting. Finally, we evaluate the impact of imputing the missing modalities. Figure~\ref{fig:overview} shows the diagram of our complete multi-modal system that comprises the following modules: 1) a feature extraction module that processes each modality individually; 2) a MIL block for each imaging modality that selects the most appropriate scan/WSI for each patient to be used in the multi-modal stage; 3) a feature reconstruction step to recover missing modalities; 4) a data fusion block; and 5) a classifier for survival prediction.
Below we discuss each of these modules and their specifications.

\subsection{Feature Extraction}
For the clinical and genomics modalities, we adopt the standard one-hot or ordinal encoding, depending on the data characteristics. In detail, all categorical variables (genomics and patient metadata) were encoded using the one-hot strategy, while T, N, M, Mp, and the tumour's grade and stage were represented using ordinal encodings. 

To characterize the WSIs, we adopted CLAM~\cite{CLAM}. This method starts by segmenting the WSIs in order to find the tissue areas and isolate them from the background. Then, the slide is partitioned into $256\times256$ patches that are forwarded to a ResNet50 for feature extraction, resulting in a set of 2,048-dimensional feature vectors. Since each WSI leads to a different number of patches and corresponding features, we perform global average pooling across the patches to compute a representation for the whole slide. This leads to a single feature vector of length 2,048.    
CT and MRI data are encoded using MeD-3D~\cite{3D-ResNet}. This is a 3D-Resnet18 model that was pretrained on 23 medical datasets. To match the requirements of MeD-3D, each scan was resized to $448\times448\times56$ and normalized as proposed in~\cite{3D-ResNet}. We extract the  $56\times56\times7\times512$ representations, which result from the application of the last convolutional layer. Then, we apply global average pooling across the 3 spatial dimensions, resulting in a feature vector of 512 dimensions.

\subsection{Imaging Data Selection using MIL }
\label{mil}
As discussed in Section~\ref{sec:img_set}, several patients present more than one CT/MRI scan and WSI image. While all this information could be used and will be made available with the dataset, we opted to reduce the amount of data used by our multi-modal system to a single CT, MRI, and WSI per patient. Since we do not know \textit{a priori} which is the best scan and WSI to characterize each patient, we implemented a novel patient-level MIL framework that automatically selects the best imaging modalities from the available pool.

We train three MIL systems, one for each modality, but they all follow the same structure. A patient is assumed to be a bag containing several instances (the features of the CT/MRI scans or WSI images) and labeled according to whether he/she survived after 12-months. Each bag will have a different number of instances, depending on the amount of available data available. All MIL systems are multilayer perceptrons trained to predict the 12-month survival. The selected MIL pooling operator is max-pooling, since it allows the selection of the most suitable scan/WSI.

The best architectures of each MIL system are the following: 1) two hidden layers with 256 and 128 units for CT and MRI; and ii) three hidden layers (512, 256, and 128) for WSI. The MIL systems are trained to predict the 12-month survival using a weighted binary cross entropy and minority class (death at 12-months) oversample ($8\times$ for CT and WSI and $16\times$ for MRI), to handle class imbalance. Finally online data augmentation was incorporated during training, by introducing Gaussian noise into the feature vectors of each modality (with $\mu=0$, $\sigma=0.01$).

During the inference phase of the system, the MIL modules are only applied if more than one scan/WSI are available for the patient. In that case, MIL uses the max-pooling operator to select the scan/WSI that exhibits the highest probability for survival at 12-months. The selected CT and MRI scans and WSI are then fed to the feature reconstruction and classification blocks.

\subsection{Reconstruction of Missing Modalities}
\label{sec:missing}

Missing data is possibly the most difficult challenge to overcome in a multi-modal setting~\cite{missingdata}. While some early and late fusion approaches are somewhat robust to missing data, approaches based on intermediate fusion will not work, limiting them to handle only data from patients with all modalities \cite{esteva2022,taleb2022contig}. This severely limits a model, as in the real world one or more modalities will be missing. 

Recently, transformer-based methods have emerged as a possibility to handle missing modalities \cite{transf-missing}. Here, modalities are treated as tokens that form a sequence of variable length. While this seems a promising direction, both the scalability and interpretability of transformer-based approaches may pose a barrier to their adoption. Other works try to explore possible synergies between modalities, by developing generative models \cite{cui2022} that can infer the missing modalities.

\begin{figure}[t]
  \centering
  \includegraphics[width=.75\linewidth]{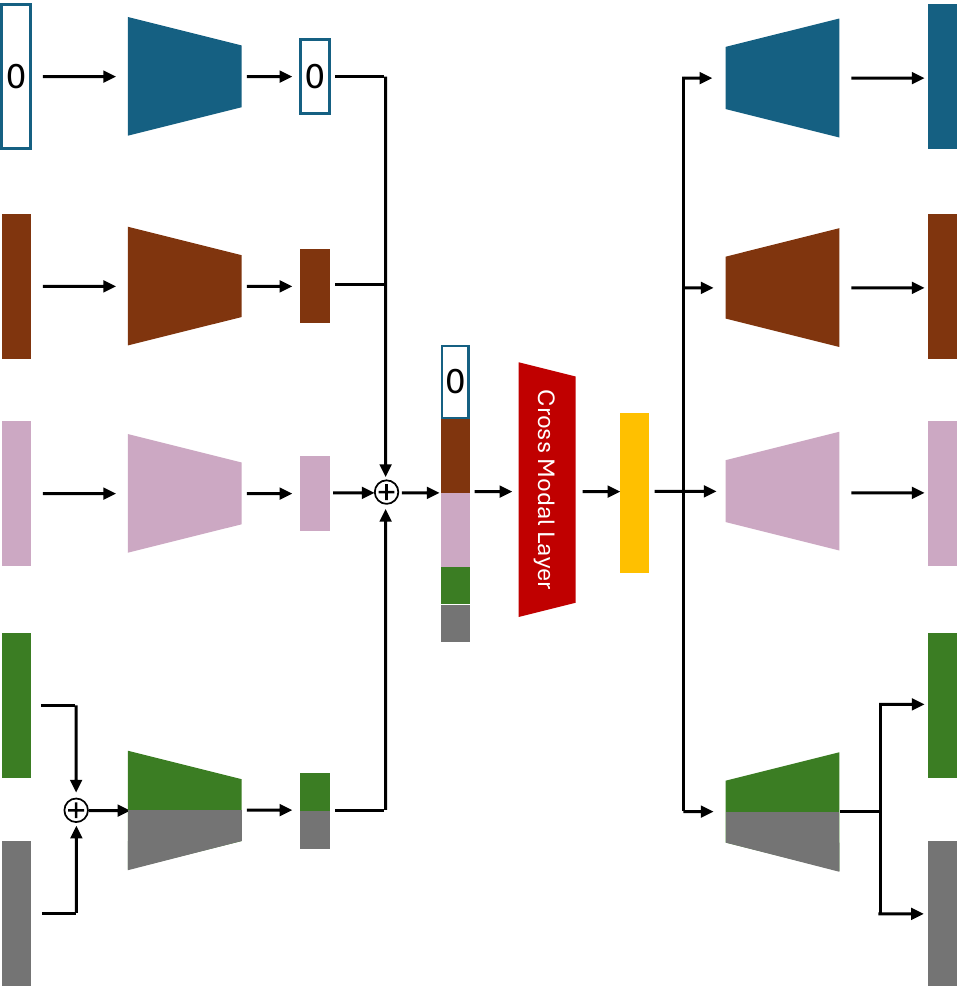}
  \caption{Reconstruction block. Each modality is processed by a specific encoder, where the missing modality vector is replaced by zeros. The concatenated ($\bigoplus$) outputs are merged in a cross modal layer, whose output (yellow vector) is fed to modality-specific decoders for reconstruction. Modality color scheme matches Fig. \ref{fig:overview}.}
  \label{fig:missing_architecture}
\end{figure}

To address the issue of missing data, we employ an encoder-decoder architecture to reconstruct the latent representations of the missing modalities, inspired by \cite{cui2022}. Figure \ref{fig:missing_architecture} shows the pipeline of our feature reconstruction method.

The encoder separately receives the feature vectors of each modality and projects them to lower dimensional spaces, using a set of MLPs. In the case where one or more modalities are missing, their feature vectors are replaced by 0s. The resulting embeddings are then concatenated and fed to a last fully connected layer that captures cross modal relationships between the different data sources. 

The decoder receives the output of the cross modal fully connected layer and feeds it to four MLPs, one to recover the feature vector of each modality. These MLPs mirror the ones used in the encoder. In the end, we obtain four feature vectors, one for each modality. In cases where modalities are missing, we expect that the decoder is able to generate a feature vector by learning cross modal relationships.

The architecture for all MLP blocks is composed of two fully connected layers with 128 neurons. The cross modal fully connected layer also comprises 128 neurons. We train the reconstruction model using the mean squared error loss with $6\times$ oversampling of the minority (death at 12 months) class. As before, we perform online data augmentation by applying Gaussian noise to the feature vectors ($\mu=0$, $\sigma=0.01$). Finally, to further improve the generalization of the system we perform modality dropout during training in cases where at most one of the modalities is missing.

During inference, we only apply the reconstruction module to patients that have missing modalities. The module receives the feature vectors of the available modalities, while the one(s) missing are replaced by vectors of 0s. The decoder outputs the feature vectors of all modalities, by reconstructing the ones available and generating the missing ones. Only the newly generated vectors are kept, as in the classification stage we will want to preserve the (correct) existing information.

\subsection{Multi-modal Data Fusion and Classification}
\label{datafusion}
In this section we will describe the design of the final classifier to predict the 12-month survival, as well as the various late and early fusion strategies that were investigated to combine the different modalities.

\subsubsection{Classifier}
\label{sec:class}
The classifier adopted in all the experiments is an MLP.  In the first set of experiments, we build a baseline classifier for each modality, in order to determine their individual predictive power. These classifiers were all trained using a weighted binary cross entropy, to account for class imbalance, as well as oversampling of the minority class ($6\times$). As in the previous sections, online data augmentation is performed by adding Gaussian noise to the feature vectors. 

When we explore the integration of the various modalities, the classifier is defined by the fusion strategy. In the case of late fusion we use the previously described classifiers and combine their outputs, as will be described in Section \ref{sec:late}. In the case of early fusion, we start by combining the feature vectors of the various modalities and then feed an MLP with this information. These MLPs are trained in a similar fashion to those of the individual modalities. The early fusion strategies are described in Section \ref{sec:early}. 

\subsubsection{Late Fusion}
\label{sec:late}
We compared two late-fusion approaches. The first one is a standard weighted sum (WS) of the outputs for the individual classifiers. The weights are computed based on the balanced accuracy (BAcc) -- the average of recalls of each classifier, and the survival prediction is given by
\begin{equation}
   P_{\text{survival}} = \frac{\sum_{m = 1}^{M} ~\mathbbm{1}_{m} ~w_m  ~p_m}{\sum_{m = 1}^{M} \mathbbm{1}_{m} ~w_m},
   \label{eq:weightedsum}
\end{equation}
where $m$ represent the modalities, $\mathbbm{1}_{m}$ is the indicator function (1 if modality exists, 0 otherwise), and $w_m$ is the BAcc of modality $m$, and $p_m$ is the corresponding survival prediction probability. This approach is agnostic to the number of available modalities, as we can just discard the missing ones if the reconstruction block is not used.

The second approach uses learned weights (LW), where each individual modality classifier is associated with a learnable weight that undergoes updates during the training process through gradient descent. The subsequent procedure closely resembles the one outlined in the weighted sum explanation and the final probability is computed using \eqref{eq:weightedsum}. However, instead of employing BAcc, the learned weights are used, and the entire weighted sum process takes place during the forward pass of the model. In the case where we do not perform feature reconstruction, we exclude the missing modalities by applying a masking operator. 

\subsubsection{Early-fusion}
\label{sec:early}
For early-fusion, two different approaches are tested: i) the concatenation of all modality features; and ii) the combination of the features through a mean operator. Since the original feature vectors of the various modalities have different shapes, we start by projecting them into the same dimension using a set of MLPs. Then we proceed with the early fusion. In the scenario where one or more modalities are missing and we do not perform feature reconstruction, we either replace the corresponding features by a vector of zeros (concatenation) or exclude the modality (masked mean). Finally, the resulting ouput is fed to the classifier. 
\begin{table*}[t]
\centering
{\small
\begin{tabular}{@{}lccccc@{}}
\toprule
Experiment     & CT       & MRI & WSI & ClinGen & All Patients \\ \midrule 
MIL  & 61.57      & 83.33  & 69.94  & - & - \\ \midrule 
Baselines       & 61.59      & 50.00  & 53.42  & 73.16 & - \\\midrule 
Late Fusion WS  & 65.00 & 73.33 & 73.16 & 73.16 & 73.15 \\
Late Fusion LW  & 73.85 & 61.66 & 69.36 & 69.36 & 69.35 \\\midrule 
Early Fusion Mean  & 88.18 & 73.33 & 83.23 & 83.23 & 82.32 \\
Early Fusion Cat  & 77.50 & \textbf{85.00} & 78.45 & 78.45 & 78.45 \\\midrule 
Early Fusion Mean (W/ Reconstruction) & \textbf{90.91} & 61.67 & \textbf{84.70} & \textbf{84.70 }& \textbf{84.70} \\
Early Fusion Cat (W/ Reconstruction) & 85.45 & 56.67 & 83.27 & 83.27 & 83.27 \\ \bottomrule
\end{tabular}}
\caption{Test BAcc results for our experiments. The All Patients column is the BAcc for the entire test set. Other columns were evaluated only on the patients with those modalities. ClinGen is the combination of clinical and genomics data. \textbf{Bold} highlights the best results.}\label{tab:res}
\end{table*}

\section{Benchmark Results}
\label{sec:exp_res}
In this section we will present and discuss the benchmarking results for MMIST-ccRCC in the task of 12-month survival prediction. The train and test splits used in all the experiments are the ones shown in Table \ref{tab:data_distribution} and the amount of patients that survived after 12 months is close to 88\%. Given this extreme imbalance, we opted to evalute the results in terms of BAcc, \textit{i.e.}, the average Recall for both classes.

The results section is structured into four main parts: i) Imaging Data Selection using MIL, where we discuss the validity of using a MIL block that selects the most suitable CT/MRI scan and WSI for each patient, to be later used in the baseline and multi-modal stages; ii) Baseline Models, where we compare the performance of individual modalities; iii) Multi-modal Data Fusion, where we compare both late and early fusion approaches; and iv) Reconstruction of Missing Modalities, where we provide an interpretable analysis of the reconstructed features and demonstrate the benefits of reconstructing missing modalities.

\subsection{Imaging Data Selection using MIL}
 The results for the patient-level MIL framework are shown in the first row of Table \ref{tab:res}. The goal of these blocks is to select the most suitable imagining modalities from the available pool (recall Section \ref{sec:img_set}) to be used in the fusion stage.

While the predictions given by the MIL models will not be used in the later stages of our system, their results provide valuable preliminary insights into the challenges that MMIST-ccRCC presents, as well as the expected performance for each modality alone. All systems exhibited a lower performance for the minority class (death at 12 months), justifying the low BAcc scores, probably due to i) the reduced number of patients in this class; and ii) the small amount of CT and MRI scans. Surprisingly, the MIL model that performs the best is the one for MRI, which is the modality with the smaller number of patients. Similar results are obtained for CT and WSI. The results for the latter are unexpectedly lower, even though WSIs are available for all patients and they have been recently explored with success in the identification of imaging biomarkers \cite{steyaert2023multimodal}.

\subsection{Baseline models}
This section showcases the results achieved by each individual modality in the task of 12-month survival prediction. As described in Section \ref{sec:class}, we trained a separate classifier for each modality in order to have a preliminary grasp of their predictive power. All baselines trained with imaging data only used the best CT/MRI scans and WSIs selected in by the MIL block. 

The single modality baseline results can be seen in the second row of Table \ref{tab:res}. These results reveal that the ClinGen modality, which combines clinical and genomics data (Section \ref{sec:cligen_set}) achieves the highest performance in terms of BAcc. Since this modality contains information about the tumour and key genetic mutations of ccRCC, these results are very promising. Furthermore, given that we have clinical data for all patients and genomics for over 74$\%$, we can perceive this score (73.16$\%$) as a the best single modality baseline to compare with the fusion results. 

As expected from the MIL results, imaging modalities exhibit comparatively lower performance. This may be due to the lack of data, in the case of CT and MRI, while the results for WSI are again unexpected. A possibility explanation is the type of features used, as either CLAM \cite{CLAM}) or the aggregation strategy (average pooling across all WSI patches) may not be discriminative enough and future studies should be conducted on this topic. The first two rows of Table \ref{tab:res} show that there has been a decrease in the performance of the model for MRI and WSI. This was expected since we are using fewer imaging data per patient. However, in sections \ref{sub:aa} and \ref{sub:missing} we will verify that we can significantly improve these results.

\subsection{Multi-modal Data Fusion}
\label{sub:aa}

From the third to sixth rows of Table \ref{tab:res}, we can compare the late and early-fusion methodologies discussed in Section \ref{datafusion}.
Besides the results for the entire test set, represented in the last column, we also report the performance on the subset of patients that contain each specific modality. This allows us to compare directly with each of the baselines, and determine whether combining modalities contributes in these cases.

Early-fusion methods emerge as the best performers, with the mean-vector being the best approach. We achieve a promising $82.32\%$ BAcc across all patients that compares favorably with the 73.16$\%$ originally achieved by the baseline of clinical and genomics data (ClinGen) and significantly outperforms the baseline of WSI (53.43\%). These improvements are even more notorious when we perform a modality-based analysis. Here we can see that adding more modalities helps mitigating the lack of data, as in the case of MRI where the BAcc improves from 50\% to 73.33\% (mean) and 85\% (concatenation).

Late fusion shows inferior results, mainly due to its dependency on the performance of the baselines. As discussed in the previous section, for most modalities the performance was not very high. Nonetheless, both late-fusion models exhibit notable improvements compared to baseline models for all imaging modalities. Between the two late fusion strategies, the standard weighted average (WS) seems to be more suitable than learning the weights (LS).

An interesting result is the performance of the models in patients with WSIs and ClinGen data, which is the same independently of the fusion strategy. This may be due to the fact that both modalities are present for almost all patients - in ClinGen, the genomics data is missing for 26\% of the patients, but the clinical data is available for all. Moreover, since part of the clinical data is derived from the inspection of the WSIs, there may be some redundancy between the two modalities.

\begin{figure*}[t]
  \centering
    
    
  
    \begin{tabular}{cc}
        \includegraphics[width=.35\linewidth]{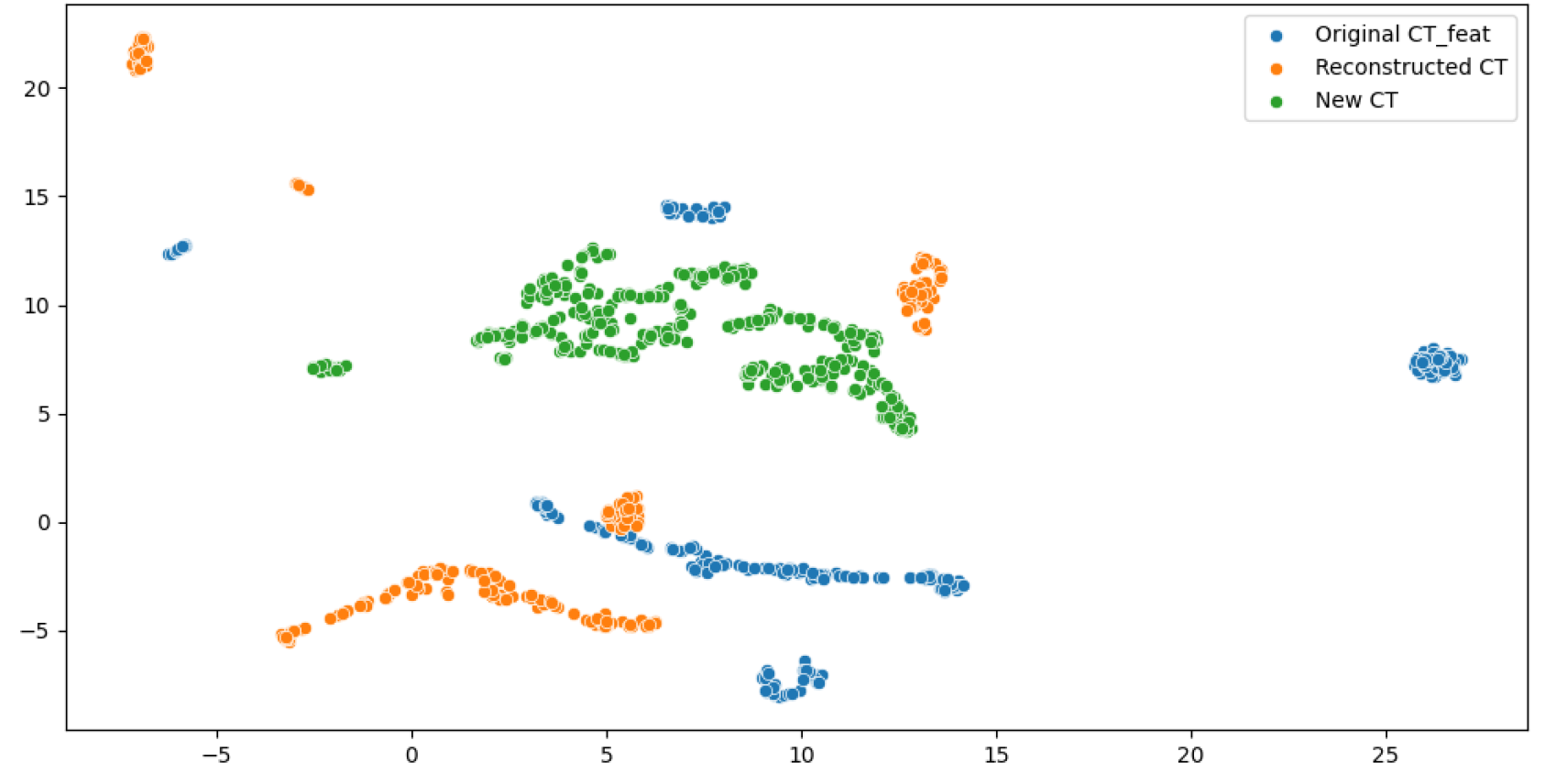} & \includegraphics[width=.35\linewidth]{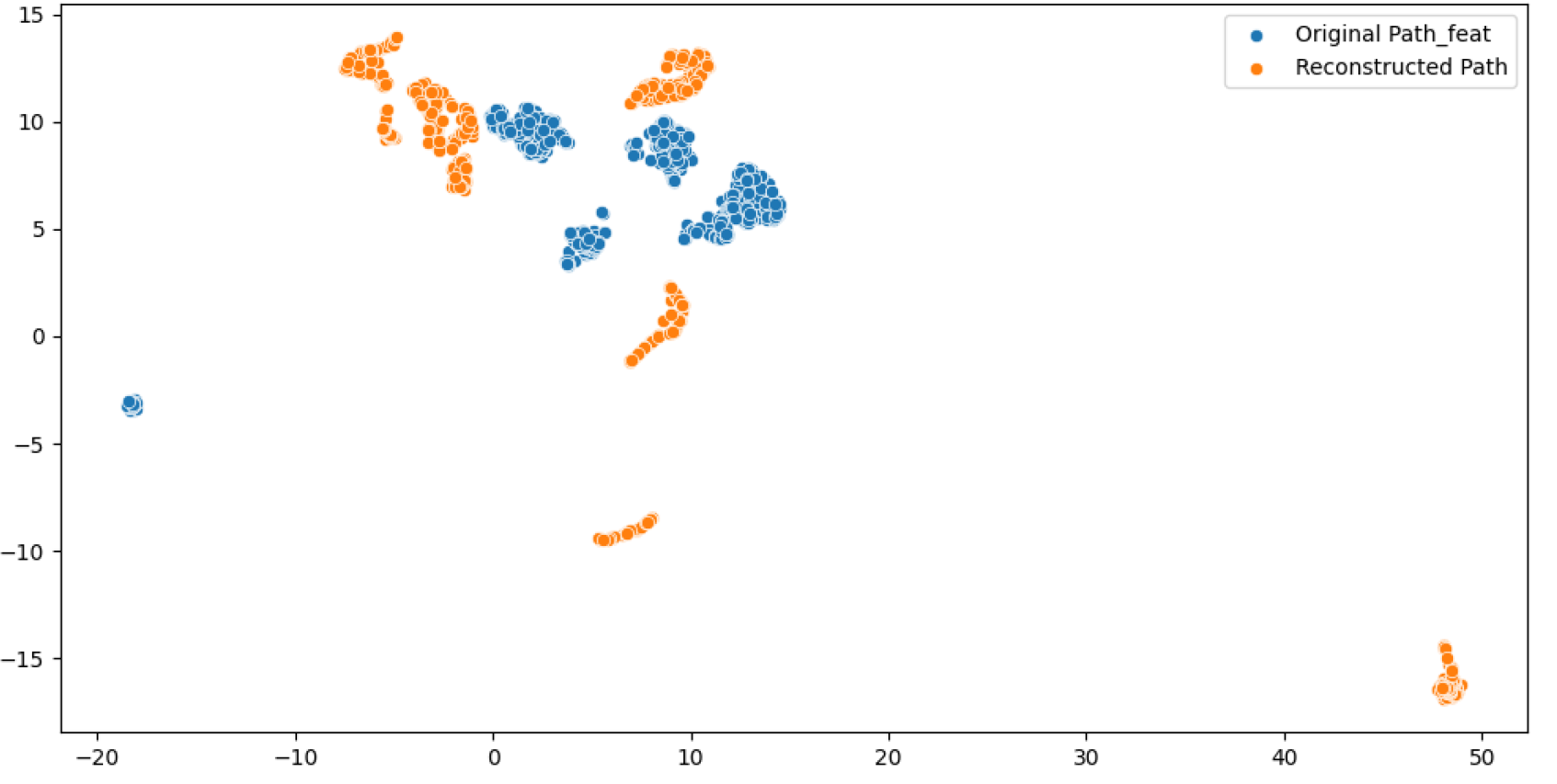} \\
        (a) CT & (b) Pathology \\
        \includegraphics[width=.35\linewidth]{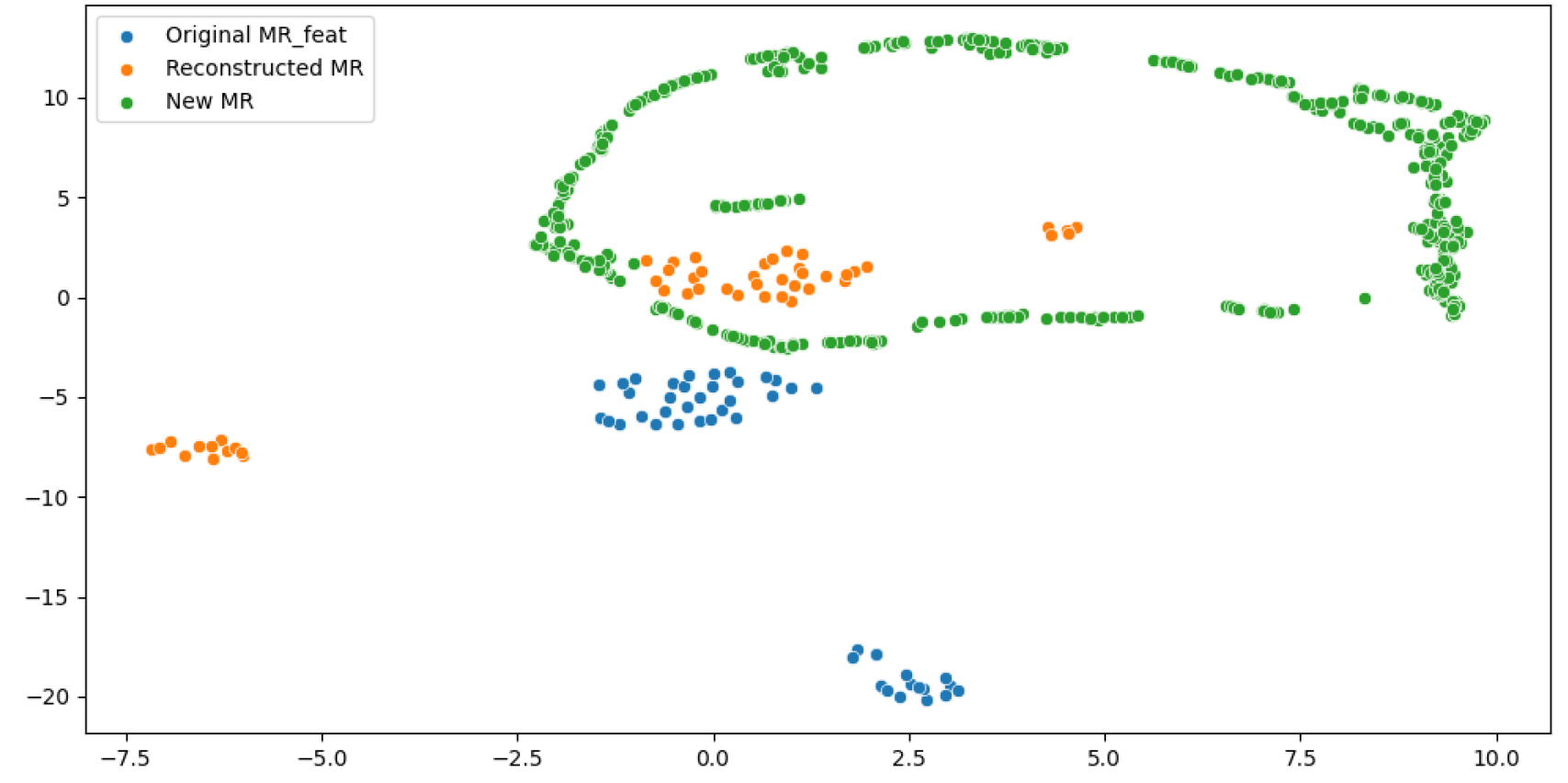} & \includegraphics[width=.35\linewidth]{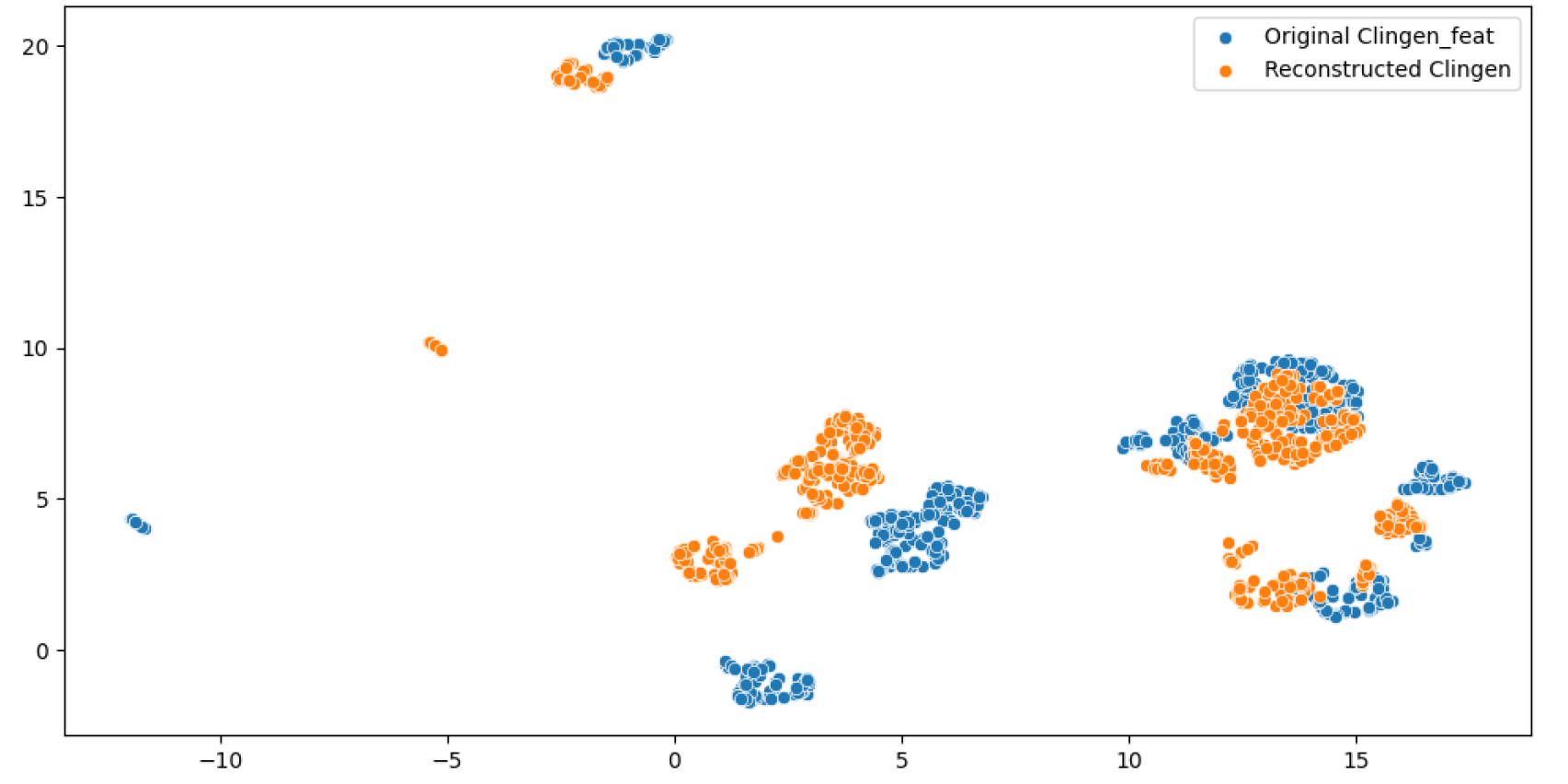} \\
        (a) MRI & (b) ClinGen \\
    \end{tabular}
  \caption{UMAP projection for all modalities for different sets of latent representations: ground truth (blue), reconstructions (orange), and newly generated feature vectors for missing modalities (green).}
  \label{fig:umap}
\end{figure*}

\subsection{Reconstruction of Missing Modalities}
\label{sub:missing}
In this section we analyze the impact of trying to recover the missing modalities using a generative approach, as described in Section \ref{sec:missing}. Our proposal generates the latent representations of the missing modalities, which can be combined with the existing ones. Below we analyze the results from the perspective of the generated features and their impact on the performance of the model.

\textbf{Interpretable Analysis:} We employed UMAP \cite{umap} to visualize high-dimensional feature reconstructions. In detail, we compare the ground truth latent representations, with the ones generated by the model (new, when the data was missing, and reconstructed otherwise). The results are depicted in Figure \ref{fig:umap}.

Comparing the original latent representations for WSI and ClinGen data with the ones obtained by the reconstruction model, it is possible to see that the clusters of ground truth and generated vectors are similar to each other and close in space. Nevertheless, it seems that the reconstruction module adds some noise to the feature vectors, as the reconstructed ones do not fully overlap with the ground truth. Since WSIs and ClinGen are available for all patients, it was expected that the reconstruction module achieve the best performances for these two modalities.

Inspecting the results for the two radiology modalities, we observe that the reconstruction module seems to work better for CT than MRI. This was expected, as the former is present for approximately 40\% of the patients, while only 8\% of the patients contain MRIs. This is further supported by inspecting the generated latent representations for the missing modalities, where for CT they seem to be located closer to the other clusters, while for MRI they are spread out.

\textbf{Quantitative Analysis:}
The last two rows of Table \ref{tab:res}  show the performance of the model after incorporating the reconstruction module. We only show results for the early fusion approaches, as these were better than the late fusion ones in the previous experiments. Nevertheless, the trend was the same. Upon the introduction of missing modality reconstruction, performance improvements are observed in both early fusion settings and for almost all modalities, particularly CT. In the case of MRI, there was a performance drop that may be justified by the quality of the newly generated latent representation (recall Figure \ref{fig:umap}). Nevertheless, we were able to achieve a final BAcc of 84.7\% (early fusion using the mean vector) that is more than 10 percentual points higher than the baseline of the ClinGen data, and superior to the BAcc achieved by the same early fusion strategy with missing data. The gains of including the generated modalities are bigger for the concatenation approach. This is somewhat expected, as in the standard concatenation we were replacing the missing modalities by vectors of 0s and now it is being replaced by the generated features.

Our findings are noteworthy considering that many patients lack certain modalities, particularly those related to radiology. They show that the reconstruction model adeptly generates pertinent information, indicating potential synergies and even some redundancy across modalities.

\subsection{Future Research Directions}
We performed various benchmarking experiments using MMIST-ccRCC and showed that our curated dataset is useful in the task of 12-month survival prediction. We have also showed that the combination of various modalities and the usage of a feature reconstruction module to generate the missing modalities can significantly improve the performance of the system against that of using a single modality for the task. Nevertheless, we are aware that MMIST-ccRCC has more potential and presents interesting lines of research for the future.

MMIST-ccRCC comprises various modalities. However, we have reduced the amount of genomics data to three key genes. Moreover, proteomics data was not included, since this information was missing for the TCGA repository. In the future, we plan to increase the presence of these two modalities in MMIST-ccRCC. Regarding the imaging modalities, we used a MIL framework to select a single CT/MRI scan and WSI per patient. However, we will release all scans and WSIs associated to each patient and encourage the community to develop strategies that take advantage of this information in their studies. Similarly, other encoding approaches for imaging data can be explored, as the results for WSI have room for improvement.

We present a set of benchmarking experiments, considering different scenarios of multi-modal systems: single modality, modality fusion, and missing modalities. While our experiments offer promising starting points, we encourage the community to investigate more complex approaches to handle each of the previous challenges. Additionally, the characteristics of MMIST-ccRCC make it possible to apply this dataset in a variety of new tasks, such as finding new imaging biomarkers or developing more advanced generative approaches for missing modalities.

\section{Conclusions}
\label{sec:conc}
This paper describes a new real world multi-modal dataset called MMIST-ccRCC that comprises more than 600 patients with ccRCC. Our curated dataset is composed of 2 radiology modalities (CT and MRI), histopathology (WSIs), clinical data, and genomics, and exhibits varying degrees of missing modalities.

We conducted various benchmarking experiments in the task of 12-month survival prediction, to showcase the relevance of our dataset. We extensively evaluate various early and late data fusion strategies, revealing that the mean operator (utilized in early fusion) consistently outperforms other methods. Moreover, we propose a novel strategy based on feature reconstruction to address the challenge of missing modalities, a common issue encountered in real-world datasets. Our findings indicate that the existing modalities contain redundant information, enabling the reconstruction model to effectively generate the missing modalities and thereby enhance the performance of survival prediction classifiers, but also complementary information as the performance increases when using multi-modal approaches.

We will keep updating MMIST-ccRCC, namely by adding proteomics and increasing the genomics profile. We expect that this work fosters future research in the multi-modal domain, and encourages the community to curate and release more datasets with similar characteristics.


\section*{Acknowledgements}

This work was supported by LARSyS funding (DOI: 10.54499/LA/P/0083/2020, 10.54499/UIDP/50009/2020, and 10.54499/UIDB/50009/2020) and projects 10.54499/2022.07849.CEECIND/CP1713/CT0001, 2023.02043.BDANA, Center for Responsible AI [PRR - C645008882-00000055].

{
    \small
    \bibliographystyle{ieeenat_fullname}
    \bibliography{refs}
}


\end{document}